\documentclass[aps,prb,amsmath,amssymb,dvips]{revtex4}

\usepackage{epsfig}
\usepackage{graphicx}
\usepackage{dcolumn}
\usepackage{bm}

\begin{document}

\title{Derivation of Lindblad master equation for the quantum Ising model interacting with a heat bath}
\author{Peizhi Mai}
\author{Shuai Yin}

\affiliation{State Key Laboratory of Optoelectronic Materials and
Technologies, School of Physics and Engineering, Sun Yat-sen
University, Guangzhou 510275, People's Republic of China}
\begin{abstract}
Starting from the Liouville-von Neumann equation, under a weak coupling limit we derive the Lindblad master equation for the one-dimensional quantum Ising model in a Markov approximation and a rotating wave approximation. We also prove that the steady solution of the Lindblad equation is the canonical distribution independent of the dissipation rate.
\end{abstract}

\pacs{}
\maketitle

\section{Introduction}
The Lindblad master equation plays an important role in open quantum
systems. As an example, it has been widely used in quantum
optics~\cite{orszag}. It was originally derived by Lindblad using quantum dynamical semigroups~\cite{Lindblad}. Attal and Joye \cite{attal} obtained it through taking a continuous limit of repeated interactions with a sequence of baths in a given density matrix state. Recently, Brasil, Fanchini, and Napolitano~\cite{brasil} provided a relatively simple derivation of the Lindblad equation starting from a general Hamiltonian and the Liouville-von Neumann equation in a Markov approximation and a rotating wave approximation under weak-coupling limit.

For many-body systems, many studies focus on the Lindblad equation for open systems interacting with the environment at their boundaries~\cite{zunkovic1,zunkovic2}. Here we derive the Lindblad master
equation for a one-dimensional quantum Ising system interacting with a heat bath at
each site following the schema given in Ref.~\onlinecite{brasil}.

Our derivation and discussion of the Lindblad master equation will proceed as follows:
In Sec.~\ref{ham} we diagonalize the Hamiltonian for a composite system consisting of a
quantum Ising chain and a heat bath and write it in the interaction picture.
In Sec.~\ref{dbme} we start from the Liouville-von Neumann equation describing the
evolution of the density matrix of the composite system to derive the
Born-Markov equation under a weak system-environment interaction limit and a Markov approximation.
These results are then combined to obtain the Lindblad equation for the quantum Ising model with the bosonic heat bath interacting at each site in Sec.~\ref{dle}.
The steady solution of Lindblad equation is shown to be the canonical distribution independent of the dissipation rate in Sec.~\ref{can}, followed by a brief summary in Sec.~\ref{sum}.

\section{\label{ham}Hamiltonian and its diagonalization}
The Hamiltonian for the quantum Ising chain interacting with a bosonic heat bath is
\begin{equation}
H=H_S+H_B+H_{SB},\label{H}
\end{equation}
with
\begin{eqnarray}
H_S=-h_x\sum_{i=1}^N\sigma_i^x-J\sum\limits_{i=1}^{N-1}\sigma_i^z\sigma_{i+1}^z,\label{HS}
\end{eqnarray}
\begin{eqnarray}
H_B=\sum_{\beta,i}\omega_\beta b_{\beta i}^\dagger b_{\beta i},\label{HB}
\end{eqnarray}
\begin{eqnarray}
H_{SB}=\sum_{\beta,i}\lambda_\beta (b_{\beta i}^\dagger+b_{\beta i})\sigma_i^x,\label{HSB}
\end{eqnarray}
where $H_S$, $H_B$, and $H_{SB}$ are the Hamiltonians for the quantum Ising chain, the heat bath, and their interaction, respectively, $N$ is the total number of spins, $\sigma_i^j$ represents the Pauli matrix along $j$ direction at site $i$, $h_x$ is a traverse field, $\lambda_\beta$ is the coupling strength, $b^\dagger_{\beta i}$ ($b_{\beta i}$) creates (annihilates) in mode $\beta$ with an energy $\omega_\beta$ a boson coupling to the spin at site $i$. Equations~(\ref{H}) to (\ref{HSB}) have been studied in Ref.~\onlinecite{patane} using Green's functions.

In order to diagonalize $H_S$, we first apply a Jordan-Wigner transformation~\cite{jordan} to $H_S$ and $H_{SB}$. After the transformation, we have
\begin{eqnarray}
H_S=&-h_x\sum_{j=1}^N(1-C_j^\dagger C_j)-J\sum_{j=1}^N(C_j^\dagger C_{j+1}^\dagger+C_j^\dagger C_{j+1}-C_j C_{j+1}^\dagger-C_jC_{j+1}),
\end{eqnarray}
\begin{eqnarray}
H_{SB}=\sum_{\beta,i}\lambda_\beta (b_{\beta i}^\dagger+b_{\beta i})(1-C_j^\dagger C_j),
\end{eqnarray}
where $C_j^\dagger$ and $C_j$ are creation and annihilation operators at site $j$ for the Jordan-Wigner fermions, respectively.

Next, a Fourier transformation leads to
\begin{eqnarray}
H_S=2J\sum_{q>0}(C_q^\dagger&C_-q)
\left(\begin{array}{cc}
\frac{h_x}{J}-\cos q&i \sin q\\
-i\sin q&-\frac{h_x}{J}+\cos q
\end{array}
\right)
\left(\begin{array}{c}
C_q\\C_{-q}^\dagger
\end{array}
\right),
\end{eqnarray}
\begin{eqnarray}
H_{SB}=\frac{-1}{\sqrt{N}}\sum_{k,q,\beta}\left[b_{\beta q}^\dagger(C_k^\dagger C_{k+q}-C_{k+q}C_k^\dagger)+b_{\beta q}(C_{k+q}^\dagger C_k-C_k C_{k+q}^\dagger)\right],
\end{eqnarray}
\begin{eqnarray}
H_B=\sum_{\beta,q}\omega_q b_{\beta q}^\dagger b_{\beta q},
\end{eqnarray}
where $C_q^\dagger$ and $C_q$ are the corresponding operators in momentum space.

Further, a Bogoliubov transformation~\cite{bogoliubov} of
\begin{eqnarray}
\eta_q=u_q^*C_q+v_{-q}C_{-q}^\dagger
\end{eqnarray}
results in
\begin{eqnarray}
H_S=\sum_q\Omega_q\eta_q^\dagger\eta_q,\label{hsb}
\end{eqnarray}
\begin{eqnarray}
H_{SB}=\sum_{j=1}^3H_{jSB},
\end{eqnarray}
with
\begin{eqnarray}
H_{jSB}=\frac {-1}{\sqrt{N}}\sum_{k,q,\beta}\lambda_\beta(b_{\beta j}^\dagger a_{jkq}+b_{\beta j}a_{jkq}^\dagger),
\end{eqnarray}
and
\begin{eqnarray}
a_{1kq}&=&(u_k^*u_{k+q}-v_{k+q}v_k^*)(\eta_k^\dagger\eta_{k+q}-\eta_{k+q}\eta_k^\dagger),\nonumber\\
a_{2kq}&=&(v_{-k}u_{k+q}-v_{k+q}u_{-k})\eta_{-k}\eta_{k+q},\nonumber\\
a_{3kq}&=&(u_k^*v_{-k-q}^*-u_{-k-q}^*v_k^*)\eta_k^\dagger\eta_{-k-q}^\dagger,\label{akq}
\end{eqnarray}
where $\Omega_q=2J\sqrt{(h_x/J)^2+1-2(h_x/J){\cos q}}$.

In the interaction picture defined by
\begin{eqnarray}
A^I(t)=\textrm{e}^{i(H_S+H_B)t}A\textrm{e}^{-i(H_S+H_B)t},\label{interaction}
\end{eqnarray}
for an operator $A$,
\begin{eqnarray}
H^I_{SB}(t)=\frac {-1}{\sqrt{N}}\sum_{j=1}^{3}\sum_{k,q,\beta}\lambda_\beta\left(b_{\beta j}^\dagger a_{jkq}e^{-i\omega_{jkq\beta}t}+b_{\beta j}a_{jkq}^\dagger e^{i\omega_{jkq\beta}t}\right),\label{18}
\end{eqnarray}
where the superscript I denotes the interaction picture and $\omega_{jkq\beta}=\omega_{jkq}-\omega_\beta$ with $\omega_{1kq}=\Omega_{k+q}-\Omega_k$, $\omega_{2kq}=\Omega_{-k}+\Omega_{k+q}$, and $\omega_{3kq}=-\Omega_k-\Omega_{-k-q}$. In Eq.~(\ref{18}), we have made use of the relations
\begin{eqnarray}
\exp(iH_St)a_{jkq}\exp(-iH_St)=a_{jkq}\exp(-i\omega_{jkq}t),\qquad \exp(iH_St)a_{jkq}^{\dagger}\exp(-iH_St)=a_{jkq}^{\dagger}\exp(i\omega_{jkq}t),\label{aeh}
\end{eqnarray}
derived from Eqs.~(\ref{hsb}) and (\ref{akq}).

\section{\label{dbme}Derivation of Born-Markov equation}
The composite system evolves according to the Liouville-von Neumann equation
\begin{eqnarray}
\frac{\partial \rho}{\partial t}=-i[H,\rho],\label{LvN}
\end{eqnarray}
where $\rho$ is the density matrix of the composite system. In the interaction picture, the Liouville-von Neumann equation becomes
\begin{eqnarray}
\frac{\partial \rho^I(t)}{\partial t}=-i[H^I_{SB}(t),\rho^I(t)].\label{Liouville}
\end{eqnarray}
Integrating Eq.~(\ref{Liouville}) from $0$ to $t$ and substituting the result back into it gives
\begin{eqnarray}
\frac{\partial \rho^I(t)}{\partial t}=-i[H^I_{SB}(t),\rho^I(0)]-[H^I_{SB}(t),\int\limits_0^tdt'[H^I_{SB}(t'),\rho^I(t')]],
\end{eqnarray}
where $\rho^I(0)=\rho_S^I(0)\otimes\rho_B^I(0)$.

Our goal is the evolution of $\rho^I_S(t)=\textrm{Tr}_B[\rho^I(t)]$, the density matrix operator of the system itself. So,
\begin{eqnarray}
\frac{\partial \rho^I_S(t)}{\partial t}=-i\textrm{Tr}_B[H^I_{SB}(t),\rho^I(0)]-\textrm{Tr}_B[H^I_{SB}(t),\int\limits_0^tdt'[H^I_{SB}(t'),\rho^I(t')]].\label{rsit}
\end{eqnarray}
Assuming
\begin{eqnarray}
\rho_B^I(0)=\frac{\prod_{\beta q}\textrm{exp}\left(\frac{-i\omega_\beta b_{\beta q}^\dagger b_{\beta q}}{k_BT}\right)}{\textrm{Tr}_B\left[\prod_{\beta q}{\textrm{exp}\left(\frac{-i\omega_\beta b_{\beta q}^\dagger b_{\beta q}}{k_BT}\right)}\right]},
\end{eqnarray}
for an equilibrium bath at the temperature $T$ ($k_B$ is Boltzmann's constant), one sees that  $-i\textrm{Tr}_B[H^I_{SB}(t),\rho^I(0)]=0$ because $\textrm{Tr}_B(b_{\beta q}^\dagger\rho^I_B(0))=\textrm{Tr}_B(b_{\beta q}\rho^I_B(0))=0$. Also, in Born approximation of weak system-environment interaction limit~\cite{Breuer}, we can write $\rho^I(t)=\rho_S^I(t)\otimes\rho_B^I(0)$ as the reduced density matrix for the bath changes little with time because of its huge size. So, Eq.~(\ref{rsit}) becomes
\begin{eqnarray}
\frac{\partial \rho^I_S(t)}{\partial t}=-\textrm{Tr}_B[H^I_{SB}(t),\int\limits_0^tdt'[H^I_{SB}(t'),\rho_S^I(t')\otimes\rho_B^I]],\label{11}
\end{eqnarray}
or
\begin{eqnarray}
\rho^I_S(t')-\rho^I_S(t)=-\int\limits_t^{t'}dt''{\rm Tr}_B[H^I_{SB}(t''),\int\limits_0^{t''}dt''' [H^I_{SB}(t'''),\rho_S^I(t''')\otimes\rho_B^I]],\label{12}
\end{eqnarray}
after integration. We see that the difference between $\rho^I_S(t')$ and $\rho^I_S(t)$ is of second order in $H^I_{SB}$. As a result, in the weak system-environment interaction limit, we can replace $\rho^I(t')$ in Eq.~(\ref{11}) with $\rho^I(t)$ and arrive at
\begin{eqnarray}
\frac{\partial \rho^I_S(t)}{\partial t}=-\textrm{Tr}_B[H^I_{SB}(t),\int\limits_0^tdt'[H^I_{SB}(t'),\rho_S^I(t)\otimes\rho_B^I]].
\end{eqnarray}
In the Markov approximation, the system is memoryless. This can be obtained by changing the integrated variable to $t-t'$ and sending the upper limit of the integral to $+\infty$~\cite{Breuer}. The result is a Born-Markov equation,
\begin{eqnarray}
\frac{\partial\rho_S^I(t)}{\partial t}=-\textrm{Tr}_B[H_{SB}^I(t),\int\limits_0^{+\infty}dt'[H_{SB}^I(t-t'),\rho_S^I(t)\otimes\rho_B^I]].\label{born}
\end{eqnarray}

\section{\label{dle}Derivation of Lindblad equation}
In this section, we shall derive the Lindblad equation from the Born-Markov equation (\ref{born}) in the rotating wave approximation~\cite{Breuer}.

We first consider the contribution from $H_{1SB}$. Denoting the corresponding density matrix as $\rho^I_{1S}$, substituting Eq.~(\ref{18}) into Eq.~(\ref{born}), and using~\cite{orszag}
\begin{eqnarray}
\textrm{Tr}_B(b_{\beta q}b^\dagger_{\beta' q'}\rho^I_B)&=&(\langle n(\omega_\beta)\rangle+1)\delta_{\beta\beta'}\delta_{qq'},\nonumber\\
\ \textrm{Tr}_B(b^\dagger_{\beta' q'}b_{\beta q}\rho^I_B)&=&\langle n(\omega_\beta)\rangle\delta_{\beta\beta'}\delta_{qq'},\nonumber\\
\textrm{Tr}_B(b_{\beta q}b_{\beta' q'}\rho^I_B)&=&\textrm{Tr}_B(b^\dagger_{\beta q}b^\dagger_{\beta' q'}\rho^I_B)=0,
\end{eqnarray}
with $\langle n(\omega_\beta)\rangle=\langle b^\dagger_{\beta q}b_{\beta q} \rangle=1/[\textrm{exp}(\omega_\beta/k_BT)-1]$, we have
\begin{eqnarray}
\frac{\partial\rho^I_{1S}}{\partial t}&=&\frac{-1}{N}\int\limits_0^{+\infty}dt'\sum_{kk'q\beta}\lambda^2_\beta\nonumber
\\&\times&\left\{\left[a^\dagger_{1kq}a_{1k'q}\rho_{1S}^I (\langle n(\omega_\beta)\rangle+1)\textrm{e}^{i\omega_{1kq\beta}t-i\omega_{1k'q\beta}(t-t')}+a_{1kq}a^\dagger_{1k'q}\rho_{1S}^I \langle n(\omega_\beta)\rangle\textrm{e}^{-i\omega_{1kq\beta}t+i\omega_{1k'q\beta}(t-t')}\right]\right.\nonumber
\\&-&\left[a^\dagger_{1kq}\rho_{1S}^Ia_{1k'q}\langle n(\omega_\beta)\rangle\textrm{e}^{i\omega_{1kq\beta}t-i\omega_{1k'q\beta}(t-t')}+a_{1kq}\rho_{1S}^Ia^\dagger_{1k'q}(\langle n(\omega_\beta)\rangle+1)\textrm{e}^{-i\omega_{1kq\beta}t+i\omega_{1k'q\beta}(t-t')}\right]\nonumber
\\&-&\left[a^\dagger_{1k'q}\rho_{1S}^Ia_{1kq}\langle n(\omega_\beta)\rangle\textrm{e}^{-i\omega_{1kq\beta}t+i\omega_{1k'q\beta}(t-t')}+a_{1k'q}\rho_{1S}^Ia^\dagger_{1kq}(\langle n(\omega_\beta)\rangle+1)\textrm{e}^{i\omega_{1kq\beta}t-i\omega_{1k'q\beta}(t-t')}\right]\nonumber
\\&+&\left.\left[\rho_{1S}^Ia^\dagger_{1k'q}a_{1kq}(\langle n(\omega_\beta)\rangle+1)\textrm{e}^{-i\omega_{1kq\beta}t+i\omega_{1k'q\beta}(t-t')}+\rho_{1S}^Ia_{1k'q}a^\dagger_{1kq}\langle n(\omega_\beta)\rangle\textrm{e}^{i\omega_{1kq\beta}t-i\omega_{1k'q\beta}(t-t')}\right]\right\}.\label{long}
\end{eqnarray}

To proceed, note first that the integral
\begin{eqnarray}
\int\limits_0^\infty dt'\textrm{e}^{\pm i\omega_{1k'q\beta}t'}=\pi\delta(\omega_{1k'q\beta})\pm{i\cal P}\frac{1}{\omega_{1k'q\beta}}
\end{eqnarray}
indicates that only the values of $k'$, $q$, and $\beta$ satisfying $\omega_{1k'q\beta}=0$ contribute to the real part of the righthand side of (\ref{long}) due to the $\delta$ function. Further, after the integration, one is left with exponentials of $\textrm{e}^{i(\omega_{1k'q\beta}-\omega_{1kq\beta})t}$, which, in the rotating wave approximation\cite{Breuer}, vanishes unless $\omega_{1k'q\beta}=\omega_{1kq\beta}$ or $k=k'$. So, Eq.~(\ref{long}) becomes
\begin{eqnarray}
\begin{split}
\frac{\partial\rho_{1S}^I}{\partial t}=&-i[H_{1LS},\rho_{1S}^I]
-c\sum_{kq}\left[\gamma_{1kq}(\langle n(\omega_{1kq})\rangle+1)(\rho_S^I a_{1kq}^\dagger a_{1kq}+a_{1kq}^\dagger a_{1kq}\rho_S^I-2a_{1kq}\rho_S^I a_{1kq}^\dagger)\right]\\
&-c\sum_{k,q}\left[\gamma_{1kq}\langle n(\omega_{1kq})\rangle(\rho_S^I a_{1kq}a_{1kq}^\dagger+a_{1kq}a_{1kq}^\dagger \rho_S^I-2a_{1kq}^\dagger \rho_Sa_{1kq})\right],\label{b}
\end{split}
\end{eqnarray}
where $H_{1LS}=(1/N)\sum_{kq}(D_{1kq}[a^\dagger_{1kq},a_{1kq}]+ \bigtriangleup\omega_{1kq}a^\dagger_{1kq}a_{1kq})$ with $D_{1kq}=\sum_\omega {\cal P}c_{\omega}\langle n(\omega)\rangle/(\omega_{1kq}-\omega)$, $\bigtriangleup\omega_{1kq}=\sum_\omega {\cal P}c_{\omega}/(\omega_{1kq}-\omega)$, and $c_{\omega}=\sum_\beta \lambda_\beta^2\delta_{\omega,\omega_\beta}$, $\gamma_{1kq}=\pi c_{\omega_{1kq}}/(cN)$, and $c=\sum_\omega c_{\omega}/N'$ ($N'$ is the total number of boson states), which is a dissipation rate describing how fast the system dissipates to equilibrium.

Similarly, the term containing only $H^I_{2SB}$ contributes an equation similar to~(\ref{b}) because of the modes satisfying $\omega_{2kq\beta}=\omega_{2k'q\beta}$ and again $k=k'$. The cross terms from $H^I_{1SB}$ and $H^I_{2SB}$ should vanish in the same approximation because the combination of $k'$, $k$ and $q$ to meet $\omega_{1kq\beta}=\omega_{2k'q\beta}$ or $\omega_{2kq\beta}=\omega_{1k'q\beta}$, if any, is much less than that satisfying either $\omega_{1kq\beta}=\omega_{1k'q\beta}$ or $\omega_{2kq\beta}=\omega_{2k'q\beta}$. Terms containing $H^I_{3SB}$ do not contribute to the real part of the righthand side of Eq.~(\ref{long}) because $\omega_{3kq\beta}$ is always negative.

Collecting the relevant terms, we then obtain the Lindblad master equation in the interaction picture
\begin{eqnarray}
\begin{split}
\frac{\partial \rho_S^I}{\partial t}=&
-c\sum_{j=1}^2\sum_{k,q}\left[\gamma_{jkq}(\langle n(\omega_{jkq})\rangle+1)(\rho_S^I a_{jkq}^\dagger a_{jkq}+a_{jkq}^\dagger a_{jkq}\rho_S^I-2a_{jkq}\rho_S^I a_{jkq}^\dagger)\right]\\
&-c\sum_{j=1}^2\sum_{k,q}\left[\gamma_{jkq}(\langle n(\omega_{jkq})\rangle(\rho_S^I a_{jkq}a_{jkq}^\dagger+a_{jkq}a_{jkq}^\dagger \rho_S^I-2a_{jkq}^\dagger \rho_Sa_{jkq})\right],\label{lidblad}
\end{split}
\end{eqnarray}
where $k$ and $q$ take values satisfying $\omega_{jkq\beta}=0$. In Eq.~(\ref{lidblad}), we have neglected a Lamb shift term $-i[H_{LS},\rho]$ with $H_{LS}=H_{1LS}+H_{2LS}+H_{3LS}$ because it is of higher order~\cite{Breuer}. Transforming back to the Schr\"{o}dinger picture, we find
\begin{eqnarray}
\begin{split}
\frac{\partial \rho_S}{\partial t}=&-i[H_S,\rho_S]
-c\sum_{j=1}^2\sum_{k,q}\left[\gamma_{jkq}(\langle n(\omega_{jkq})\rangle+1)(\rho_S a_{jkq}^\dagger a_{jkq}+a_{jkq}^\dagger a_{jkq}\rho_S-2a_{jkq}\rho_S a_{jkq}^\dagger)\right]\\
&-c\sum_{j=1}^2\sum_{k,q}\left[\gamma_{jkq}(\langle n(\omega_{jkq})\rangle(\rho_Sa_{jkq}a_{jkq}^\dagger+a_{jkq}a_{jkq}^\dagger \rho_S-2a_{jkq}^\dagger \rho_Sa_{jkq})\right].\label{lindblad}
\end{split}
\end{eqnarray}

In order to write Eq.~(\ref{lindblad}) in the familiar Lindblad
form, for two energy levels $E_l$ and $E_m$ with
$E_m-E_l=\omega_{jkq}$, we may let $V_{m\rightarrow l}=a_{jkq}$ and
$V_{m\rightarrow l}^\dagger=a^\dagger_{jkq}$, which are respectively
thermal jump operators representing emitting and absorbing a particle
with energy $\omega_{jkq}$ and jumping to a lower and higher energy
state. Also, let $W_{l\rightarrow m}=\gamma_{jkq}\langle
n(\omega_{jkq})\rangle$ and $W_{m\rightarrow l}=\gamma_{jkq}(\langle
n(\omega_{jkq})\rangle+1)$, which are transition probabilities from
$l$th state to the $m$th and vice versa, so that~\cite{Binder}
\begin{eqnarray}
\frac{W_{l\rightarrow m}}{W_{m\rightarrow l}}=\frac{\gamma_{jkq}(\langle
n(\omega_{jkq})\rangle)}{\gamma_{jkq}(\langle
n(\omega_{jkq})\rangle+1)}=\textrm{exp}\left(\frac{-\omega_{jkq}}{k_BT}\right)\equiv
\textrm{exp}\left(\frac{E_l-E_m}{k_BT}\right).\label{c}
\end{eqnarray}
Equation~(\ref{lindblad}) then becomes
\begin{eqnarray}
\begin{split}
\frac{\partial \rho_S}{\partial
t}&=-i[H_S,\rho_S]-c\sum_{m>l}W_{l\rightarrow m}(V_{l\rightarrow
m}^\dagger V_{l\rightarrow m} \rho_S+\rho_S V_{l\rightarrow
m}^\dagger V_{l\rightarrow m}-2V_{l\rightarrow m} \rho_S
V_{l\rightarrow m}^\dagger)\\&-c\sum_{m>l}W_{m\rightarrow l}(V_{l\rightarrow m} V_{l\rightarrow m}^\dagger \rho_S+\rho_S
V_{l\rightarrow m} V_{l\rightarrow m}^\dagger-2V_{l\rightarrow
m}^\dagger \rho_S V_{l\rightarrow m})
\\&=-i[H_S,\rho_S]-c\sum_{l,m,m\neq l}W_{l\rightarrow m}(V_{l\rightarrow m}^\dagger V_{l\rightarrow m} \rho_S+\rho_S V_{l\rightarrow m}^\dagger V_{l\rightarrow m}-2V_{l\rightarrow m} \rho_S V_{l\rightarrow m}^\dagger),\label{21}
\end{split}
\end{eqnarray}
the usual Lindblad form. Although the form of the transition probability
$W_{l\rightarrow m}$ depends on the environment and determines the details of the process, universal
properties only rely on $W_{l\rightarrow m}/W_{m\rightarrow l}$~\cite{Binder}. For example, in Ref.~(12), $W_{l\rightarrow
m}=\beta_m$ with $\beta_m$ the probability for the system to stay in
the $m$th state in equilibrium. This completes our derivation of the
Lindblad equation for the quantum Ising model.

\section{\label{can}Canonical distribution is the steady solution of the Lindblad equation}
In this section, we discuss the steady solution of Lindblad equation.

The meaning of the Lindblad equation (\ref{21}) is clear.
If the second term in the right hand side is neglected, it is the quantum Liouville equation determining
the quantum fluctuations of the evolution of density operator $\rho_S$; while if the first term on the right hand side is neglected, the diagonal part is
\begin{equation}
\frac{\partial \rho_{ii}}{\partial t}=c\sum_{j\neq
i}(W_{j\rightarrow i} \rho_{jj}-W_{i\rightarrow j} \rho_{ii})
\label{cmaster},
\end{equation}
which is the classical master equation, and the off-diagonal part is
\begin{equation}
\frac{\partial \rho_{ij}}{\partial t}=-\frac{c}{2}(\sum_{k\neq
i}W_{i\rightarrow k}+\sum_{l\neq
j}W_{j\rightarrow l})\rho_{ij} \label{qdeco},
\end{equation}
which decays exponentially. Thus the Lindblad equation (\ref{21}) naturally integrates the quantum and thermal fluctuations together.

It can be readily checked that for a time-independent Hamiltonian, the equilibrium density matrix
of the canonical distribution $\rho_E=\textrm{exp}(-H_S/k_BT)/\textrm{Tr}[\textrm{exp}(-H_S/k_BT)]$ is
the steady solution of the Lindblad equation~\cite{attal}.
To this end, we note first that there is a detailed balance condition,
\begin{eqnarray}
\begin{split}
W_{l\rightarrow m} \rho_EV_{m\rightarrow l}
&=\gamma_{jkq}\langle n(\omega_{jkq})\rangle\frac{\textrm{exp}(-H_S/k_BT)}{\textrm{Tr}\left[\textrm{exp}(-H_S/k_BT)\right]} a_{jkq}\textrm{exp}\left(\frac{H_S}{k_BT}\right)\textrm{exp}\left(-\frac{H_S}{k_BT}\right)
\\&=\gamma_{jkq}\langle n(\omega_{jkq})\rangle\textrm{exp}\left(\frac{\omega_{jkq}}{k_BT}\right) a_{jkq}\frac{\textrm{exp}(-H_S/k_BT)}{\textrm{Tr}\left[\textrm{exp}(-H_S/k_BT)\right]}
\\&=W_{m\rightarrow l}V_{m\rightarrow l}\rho_E,\label{detail}
\end{split}
\end{eqnarray}
where uses have been made of Eqs.~(\ref{aeh}) and (\ref{c}).
Then, on the righthand side of Eq.~(\ref{21}), the first term
$[H_S,\rho_E]=0$. For the second term, using the property of the thermal jump matrices
$V_{l\rightarrow m}=V_{m\rightarrow l}^\dagger$
from their definitions and substituting Eq. (\ref{detail}) into the righthand side of the Lindblad
equation, the second term can be explicitly written as~\cite{attal},
\begin{eqnarray}
W_{l\rightarrow m}(V_{l\rightarrow m}^\dagger V_{l\rightarrow m}
\rho_E+\rho_E V_{l\rightarrow m}^\dagger V_{l\rightarrow m}-2V_{l\rightarrow m} \rho_E V_{l\rightarrow m}^\dagger)
=W_{l\rightarrow m} V_{l\rightarrow m}^\dagger V_{l\rightarrow m}
\rho_E-W_{l\rightarrow m} V_{l\rightarrow m}V_{l\rightarrow m}^\dagger\rho_E,
\end{eqnarray}
and
\begin{eqnarray}
W_{m\rightarrow l}(V_{m\rightarrow l}^\dagger V_{m\rightarrow l}
\rho_E+\rho_E V_{m\rightarrow l}^\dagger V_{m\rightarrow l}-2V_{m\rightarrow l} \rho_E V_{m\rightarrow l}^\dagger)
=W_{m\rightarrow l} V_{m\rightarrow l}^\dagger V_{m\rightarrow l}
\rho_E-W_{m\rightarrow l} V_{m\rightarrow l}V_{m\rightarrow l}^\dagger\rho_E.
\end{eqnarray}
These two terms cancel with each other and thus the second term of the
right hand side of the Lindblad equation~(\ref{21}) equals zero too. This derivation is similar to that from Ref.~(3), but we take the thermal jump involving all energy levels into account. Thus
for the weak coupling situation, the Lindblad equation reduces to the canonical
distribution solution in long time. Note that
this steady solution does \emph{not} depends on $c$.

\section{\label{sum}Summary}
We have derived the Lindblad equation for the quantum Ising
chain weakly interacting with a heat bath. Further we have confirmed that
the steady solution of this equation is the equilibrium canonical
distribution independent of the dissipation rate.


\begin{thebibliography}{99}

\bibitem{orszag}M. Orszag, {\it Quantum Optics} (Springer-Verlag, Berlin,  Heidelberg, 2000).
\bibitem{Lindblad}G. Lindblad, Commun. Math. Phys. 48, 119 (1976).
\bibitem{attal}S. Attal and A. Joye, J. Func. Analysis. 247, 253-288 (2007).
\bibitem{brasil}C. A. Brasil, F. F. Fanchini, and R. d. J. Napolitano, arXiv:1110.2122v1 (2011).
\bibitem{zunkovic1}B. Zunkovic and T. Prosen, arXiv:1007.2922v1 (2010).
\bibitem{zunkovic2}B. Zunkovic and T. Prosen, arXiv:1203.0943v3 (2012).
\bibitem{patane}D. Patan\`{e}, A. Silva, L. Amico, R. Fazio, and G. E. Santoro, Phys. Rev. Lett. 101 175701 (2008).
\bibitem{jordan}P. Jordan and E. Wigner, Z. Phys. 47, 631 (1928).
\bibitem{bogoliubov}N. N. Bogoliubov, J. Phys. (USSR), 11:23 (1947).
\bibitem{Breuer}H. P. Breuer and F. Petruccione {\it The theory of open quantum systems} (Oxford, \emph{place}, 2003).
\bibitem{Binder}D. P. Landau and K. Binder, {\it A Guide to Monte Carlo Simulations in Statistical Physics} (Cambridge, \emph{place}, 2000).
\bibitem{Yin}S. Yin, X-Z. Qin, C-H. Lee, and F. Zhong, arXiv:1207.1602v1 (2012).
\end{thebibliography}
\end{document}